\documentclass[aps,prd,twocolumn,nofootinbib]{revtex4}

\usepackage{graphicx,color}

\usepackage[colorlinks=true, linkcolor=blue, citecolor=blue, urlcolor=blue]{hyperref}

\begin{document}

\title{Self-consistent analysis for the $\eta_c\rightarrow \gamma\gamma$ process}

\author{Sheng-Quan Wang$^{1}$}
\email[email:]{sqwang@alu.cqu.edu.cn}

\author{Zhu-Yu Ren$^{1}$}

\author{Jian-Ming Shen$^{2}$}
\email[email:]{shenjm@hnu.edu.cn}

\author{Xing-Gang Wu$^3$}
\email[email:]{wuxg@cqu.edu.cn}

\author{Leonardo Di Giustino$^{4,5}$}
\email[email:]{leonardo.digiustino@uninsubria.it}

\author{Stanley J. Brodsky$^6$}
\email[email:]{sjbth@slac.stanford.edu}

\address{$^1$Department of Physics, Guizhou Minzu University, Guiyang 550025, P.R. China}
\address{$^2$School of Physics and Electronics, Hunan University, Changsha 410082, P.R. China}
\address{$^3$Department of Physics, Chongqing University, Chongqing 401331, P.R. China}
\address{$^4$Department of Science and High Technology, University of Insubria, via Valleggio 11, I-22100, Como, Italy}
\address{$^5$INFN, Sezione di Milano-Bicocca, 20126 Milano, Italy}
\address{$^6$SLAC National Accelerator Laboratory, Stanford University, Stanford, California 94039, USA}

\date{\today}

\begin{abstract}

The next-to-next-to-leading-order (NNLO) pQCD predictions for both the decay width and the transition form factor in the $\eta_c\rightarrow \gamma\gamma$ process, based on nonrelativistic QCD (NRQCD), deviate from precise experimental measurements. These significant discrepancies have cast doubt on the applicability of NRQCD to charmonium processes. In this paper, we analyze the $\eta_c\rightarrow \gamma\gamma$ process by applying the Principle of Maximum Conformality (PMC), a systematic method for eliminating renormalization scheme and scale ambiguities. The PMC renormalization scales are determined by absorbing the non-conformal $\beta$ terms which govern the behavior of the QCD running coupling via the Renormalization Group Equation. We obtain the PMC scale $Q_\star=4.49\,m_c$ for the $\eta_c\rightarrow \gamma\gamma$ decay width. Even after using the PMC method, the convergence of the pQCD series is still poor, which indicates the importance of uncalculated NNNLO and higher-order terms. The resulting value for $\Gamma_{\eta_c\rightarrow \gamma\gamma}$ is in agreement with the Particle Data Group's reported value of $\Gamma_{\eta_c\rightarrow \gamma\gamma}=5.1\pm0.4$ keV within the bounds of uncertainties. Moreover, the transition form factor obtained using the PMC is also in good agreement with precise experimental measurements. The application of the PMC suggests a potential resolution to $\eta_c\rightarrow \gamma\gamma$ puzzle and supports the applicability of NRQCD to charmonium processes.

\end{abstract}

\maketitle

\section{Introduction}
\label{sec:1}

Heavy quarkonium physics is of central interest for the understanding of quantum chromodynamics (QCD). Since the bound-state heavy quarks are nonrelativistic, quarkonium phenomena can be analyzed using the nonrelativistic QCD (NRQCD) factorization approach~\cite{Bodwin:1994jh}. For example, the annihilation of the lowest charmonium state
$\eta_c$ into photons, and its production in
$\gamma\gamma$ collisions provide sensitive tests of the application of NRQCD.

The $\eta_c\rightarrow \gamma\gamma$ decay width $\Gamma_{\eta_c\rightarrow\gamma\gamma}$ is given by
\begin{eqnarray}
\Gamma_{\eta_c\rightarrow\gamma\gamma}=\frac{\pi}{4}\,\alpha^2\,m^3_{\eta_c}\,|F(0)|^2,
\label{EQ:F0}
\end{eqnarray}
where $\alpha$ denotes the electromagnetic coupling constant, $m_{\eta_c}$ is the $\eta_c$ meson mass, and $F(0)$ is the transition form factor at zero momentum transfer. The $\eta_c\rightarrow \gamma\gamma$ decay process has been extensively studied theoretically~\cite{Godfrey:1985xj, Czarnecki:2001zc, Bodwin:2001pt, Dudek:2006ut, Feng:2015uha, CLQCD:2016ugl, Chen:2016bpj, Feng:2017hlu, Brambilla:2018tyu, Liu:2020qfz, CLQCD:2020njc, Zhang:2021xvl, Meng:2021ecs, Li:2021ejv, Colquhoun:2023zbc, ExtendedTwistedMass:2022ofm,Abreu:2022cco}. The state-of-the-art NRQCD prediction including the next-to-next-to-leading order (NNLO) pQCD calculation gives the decay width $\Gamma_{\eta_c\rightarrow\gamma\gamma}=9.7\sim10.8$ keV (with a branching ratio of Br($\eta_c\rightarrow\gamma\gamma)=(3.1\sim3.3)\times10^{-4}$)~\cite{Feng:2017hlu}.  However, this result is more than $10\sigma$ away from the Particle Data Group's reported value of $\Gamma_{\eta_c\rightarrow \gamma\gamma}=5.1\pm0.4$ keV~\cite{ParticleDataGroup:2024cfk}. This eminent discrepancy has casted doubt on the applicability of NRQCD to charmonium processes.

Lattice QCD (LQCD) calculations for the $\eta_c\rightarrow \gamma\gamma$ decay process show strong systematic uncertainties. The first LQCD calculation of $\eta_c\rightarrow \gamma\gamma$ was presented in Ref.\cite{Dudek:2006ut}. This early result~\cite{Dudek:2006ut} and subsequent LQCD results~\cite{CLQCD:2016ugl, Liu:2020qfz, CLQCD:2020njc, ExtendedTwistedMass:2022ofm} predict $\eta_c\rightarrow \gamma\gamma$ decay widths much less than the PDG value. More recent LQCD results~\cite{Meng:2021ecs, Colquhoun:2023zbc} show larger values for the $\eta_c\rightarrow \gamma\gamma$ decay width, exceeding the PDG value.  The available LQCD results thus cannot satisfactorily explain the PDG value.

In NRQCD, the $\mathcal{O}(\alpha_sv^2)$ correction for the $\eta_c\rightarrow \gamma\gamma$ decay width appears to be phenomenologically negligible~\cite{Guo:2011tz, Jia:2011ah}. Thus, this substantial disagreement cannot be explained by taking higher Fock states into consideration. It is well known that the $\eta_c\rightarrow \gamma\gamma$ decay process suffers poor pQCD convergence, as well as large renormalization scale $\mu_r$ uncertainty.

According to conventional practice, the renormalization scale is simply chosen as $m_c$ in order to eliminate the large logarithmic terms $\ln\frac{\mu^2_r}{m^2_c}$; the scale is then varied over an arbitrary range to estimate its uncertainty for the $\eta_c\rightarrow \gamma\gamma$ decay process. This conventional procedure violates the fundamental principle of renormalization group invariance, and it is affected by renormalization scheme-and-scale ambiguities in pQCD predictions. The resulting pQCD series has the ``renormalon" $n$-factorial divergence~\cite{Beneke:1998ui}. This conventional scale-setting procedure is also inconsistent with the well-known {Gell-Mann}-Low (GM-L) method used in QED~\cite{GellMann:1954fq}. Predictions for non-Abelian QCD in the limit of $N_C\rightarrow0$~\cite{Brodsky:1997jk}, must agree analytically with the predictions for Abelian QED, including the renormalization scale-setting procedure.

For the $\eta_c\rightarrow \gamma\gamma$ decay process, one cannot decide whether the poor convergence is an intrinsic property of its pQCD series, or if it is simply due to the improper choice of the scale taken in a particular range of values. Improved analyses were given in Refs.\cite{Wang:2018lry, Yu:2020tri} for $\eta_c$ physical processes, which showed the importance of correct renormalization scale-setting. A more detailed study of the renormalization scale-setting problem is in fact an important improvement for pQCD predictions.

The Principle of Maximum Conformality (PMC)~\cite{Brodsky:2011ta, Brodsky:2012rj, Brodsky:2011ig, Mojaza:2012mf, Brodsky:2013vpa} has been proposed in order to eliminate renormalization scheme-and-scale ambiguities. The PMC predictions satisfy the requirements of the renormalization group invariance~\cite{Brodsky:2012ms, Wu:2014iba, Wu:2019mky}. The PMC method generalizes the BLM scale-setting procedure~\cite{Brodsky:1982gc} to all orders. The PMC scales are determined by absorbing the $\beta$ terms which govern the behavior of the running $\alpha_s$ via the Renormalization Group Equations (RGE), reflecting the virtuality of the propagating gluons of the QCD subprocesses. As expected, the PMC method reduces correctly in the $N_C\to 0$ Abelian limit~\cite{Brodsky:1997jk} to the well known Gell-Mann-Low method~\cite{GellMann:1954fq}.

The PMC method has been successfully applied to charmonium physical processes, such as the $\gamma\gamma^*\rightarrow\eta_c$ process~\cite{Wang:2018lry}, the ratios of the $\eta_c$ decays to light hadrons or $\gamma\gamma$~\cite{Yu:2019mce}, the $e^+e^-\rightarrow\eta_c+\gamma$ process~~\cite{Yu:2020tri} and the $\chi_{c0, c2}\rightarrow\gamma\gamma$ process~\cite{Zhou:2021zvx}. In this paper, we perform a more extensive and comprehensive analysis for the $\eta_c\rightarrow \gamma\gamma$ process at NNLO by applying the PMC scale-setting to both the decay width and the transition form factor.

\section{PMC scale-setting for the $\eta_c\rightarrow \gamma\gamma$ process}
\label{sec:2}

In the NRQCD framework, the transition form factor at zero momentum transfer $F(0)$ of Eq.(\ref{EQ:F0}), calculated at NNLO in the $\overline{\rm MS}$ scheme can be written as
\begin{eqnarray}
F(0)=c^{(0)}\left[1+\delta^{(1)}\,a_s(\mu_r)+\delta^{(2)}(\mu_r)\,a^2_s(\mu_r)\right].
\end{eqnarray}
Here, the coupling constant is $a_s(\mu_r)=\alpha_s(\mu_r)/\pi$, and $\mu_r$ stands for the renormalization scale. The leading order (LO) QCD correction is $c^{(0)}=(e^2_c\langle\eta_c|\psi^\dag\chi(\mu_\Lambda)|0\rangle)/m^{5/2}_c$, where $e_c$ is the $c$-quark electric charge, $m_c$ is the $c$-quark mass, and $\langle\eta_c|\psi^\dag\chi(\mu_\Lambda)|0\rangle$ represents nonperturbative matrix element. The next-to-leading order (NLO) QCD coefficient is $\delta^{(1)}=C_F(\pi^2/8-5/2)$, the NNLO QCD coefficient $\delta^{(2)}$ is
\begin{eqnarray}
\delta^{(2)}(\mu_r)&=&\delta^{(1)}\frac{\beta_0}{4}\ln{\frac{\mu^2_r}{m^2_c}}-\frac{\pi^2}{2}\,C_F\left(C_F+\frac{C_A}{2}\right)\ln{\frac{\mu^2_\Lambda}{m^2_c}}\nonumber\\
&&+f^{(2)}_{lbl}+f^{(2)}_{reg}.
\label{delta2}
\end{eqnarray}
Here, $C_F=4/3$, $C_A=3$, and $\mu_\Lambda$ denotes the factorization scale. The $f^{(2)}_{lbl}$ is the light-by-light contribution and the $f^{(2)}_{reg}$ stands for the regular contribution. The NNLO coefficient in the $\overline{\rm MS}$ scheme was given in Ref.\cite{Feng:2015uha}. At present NNLO level, the PMC scale is determined by absorbing the
$\beta_0=11-2/3\,n_f$ term into the QCD running coupling. It is noted that the $n_f$ term in the light-by-light contribution $f^{(2)}_{lbl}$ is free of ultraviolet divergences, which should be treated as a part of the conformal contribution when applying the PMC method. The regular contribution can be separated into two parts, i.e., $f^{(2)}_{reg}=f^{(2)}_{reg,in}+f^{(2)}_{reg,n_f}\,n_f$, where the $n_f$ term associated with ultraviolet divergences should be absorbed into the QCD coupling constant.

Once the $n_f$-terms have been properly labelled, the NNLO coefficient $\delta^{(2)}$ of Eq.(\ref{delta2}) can be further divided into $\delta^{(2)}(\mu_r)=\delta_{in}^{(2)}(\mu_r)+\delta_{n_f}^{(2)}(\mu_r)\,n_f$, where
\begin{eqnarray}
\delta_{in}^{(2)}(\mu_r)&=&\frac{11}{4}\delta^{(1)}\ln{\frac{\mu^2_r}{m^2_c}}-\frac{17}{9}\pi^2\ln{\frac{\mu^2_\Lambda}{m^2_c}}\nonumber\\
&&+f^{(2)}_{lbl}+f^{(2)}_{reg,in}, \\
\delta_{n_f}^{(2)}(\mu_r)&=&\delta^{(2)}_{reg,n_f}-\frac{1}{6}\delta^{(1)}\ln{\frac{\mu^2_r}{m^2_c}}.
\end{eqnarray}

By applying the PMC method to calculate the bound state process in the physical $V$ scheme, reliable predictions can be achieved~\cite{Yan:2023mjj}. In addition, the PMC scale in QED is identical to the QCD PMC scale in the physical $V$ scheme~\cite{Wang:2020ckr}. We report here the detailed analysis for the $\eta_c\rightarrow \gamma\gamma$ decay process by applying the PMC method together with the use of the physical $V$ scheme. The PMC predictions are scheme independent, which are ensured by the PMC conformal series, and are explicitly displayed in the form of ``commensurate scale relations" (CSR)~\cite{Brodsky:1994eh, Lu:1992nt}

The $V$ scheme is defined by the static limit of the scattering potential between two heavy quark-antiquark test charges~\cite{Appelquist:1977tw, Fischler:1977yf, Peter:1996ig, Schroder:1998vy, Smirnov:2008pn, Smirnov:2009fh, Anzai:2009tm}, i.e., $V(Q^2) = -{4\,\pi^2\,C_F\,a^V_s(Q)/Q^2}$ at the momentum transfer $q^2=-Q^2$. For the QED case, when the scale of the coupling $a^V_s$ is identified with the exchanged momentum, all vacuum polarization corrections are resummed into the coupling $a^V_s$. By using the coupling constant relation between the $\overline{\rm MS}$ scheme and the $V$ scheme at the one-loop level, i.e., $a_s = a^V_s+B_2\,\left(a^V_s\right)^2$ with $B_2=-31/12+(5/18)\,n_f$, we convert the transition form factor $F(0)$ of Eq.(\ref{EQ:F0}) from the $\overline{\rm MS}$ scheme to the $V$ scheme.
\begin{eqnarray}
F(0)&=&c^{(0)}\left[1+\delta^{(1)}_V\,a^V_s(\mu_r)+\left(\delta_{in,V}^{(2)}(\mu_r)\right.\right. \nonumber\\
&&\left.\left.+\delta_{n_f,V}^{(2)}(\mu_r)\,n_f\right)\,\left(a^V_s(\mu_r)\right)^2\right],
\end{eqnarray}
where, $\delta^{(1)}_V=\delta^{(1)}$ and the NNLO perturbative coefficients in the $V$ scheme are $\delta_{in,V}^{(2)}(\mu_r)=\delta^{(2)}_{in}(\mu_r)-{31\over12}\,\delta^{(1)}$, and $\delta_{n_f,V}^{(2)}(\mu_r)=\delta^{(2)}_{n_f}(\mu_r)+{5\over18}\,\delta^{(1)}$.
After using the PMC, we obtain
\begin{eqnarray}
F(0)&=&c^{(0)}\left[1+\delta^{(1)}_V\,a^V_s(Q_\star)+\delta_{con,V}^{(2)}(\mu_r)\,\left(a^V_s(Q_\star)\right)^2\right].  \nonumber\\
\label{PMC:F0}
\end{eqnarray}
The PMC scale $Q_\star$ is given by:
\begin{eqnarray}
Q_\star=\mu_r\exp\left[\frac{3\,\delta^{(2)}_{n_f,V}(\mu_r)}{2\,T_R\,\delta^{(1)}_V}\right],
\label{PMCscaleV}
\end{eqnarray}
and the conformal coefficient $\delta_{con,V}^{(2)}(\mu_r)$ is
\begin{eqnarray}
\delta_{con,V}^{(2)}(\mu_r)=\frac{11\,C_A\,\delta^{(2)}_{n_f,V}(\mu_r)}{4\,T_R}+\delta^{(2)}_{in,V}(\mu_r).
\end{eqnarray}
At the present level;  i.e., at NNLO, the resulting PMC scale $Q_\star$ is independent of the renormalization scale
$\mu_r$. Only the conformal coefficient remains in the expression for the transition form factor $F(0)$. The conformal coefficient is also independent of the renormalization scale $\mu_r$. Thus the PMC result for $F(0)$ of Eq.(\ref{PMC:F0}) eliminates the renormalization scale uncertainty.

\section{Numerical results and discussions}
\label{sec:3}

For the numerical calculation, we take the $c$-quark mass $m_c=1.5$ GeV, the $\eta_c$ meson mass $m_{\eta_c}=2.9839$ GeV~\cite{ParticleDataGroup:2024cfk}, the electromagnetic coupling constant $\alpha=1/132.6$~\cite{Bodwin:2007ga}. The NRQCD matrix element is $\langle\eta_c|\psi^\dag\chi(\mu_\Lambda)|0\rangle^2|_{\mu_{\Lambda}=1 {\rm GeV}}=0.437$ GeV$^3$~\cite{Bodwin:2007fz, Chung:2010vz}. By evolving the matrix element from $1$ GeV to $1.5$ GeV, we obtain $\langle\eta_c|\psi^\dag\chi(\mu_\Lambda)|0\rangle^2|_{\mu_{\Lambda}=1.5 {\rm GeV}}=0.409$ GeV$^3$ by using the one-loop evolution formulae given in Ref.\cite{Bodwin:1994jh}. We adopt the two-loop $\alpha_s$ coupling constant;  its corresponding $\Lambda^{\overline{MS}}_{\rm QCD}$ can be determined by the world average value $\alpha_s(M_Z)=0.1180$~\cite{ParticleDataGroup:2024cfk}. The asymptotic scale in the $V$ scheme is obtained by $\Lambda^{V}_{\rm QCD}=\Lambda^{\overline{MS}}_{\rm QCD}\,\exp[-2B_2/\beta_0]$.

\subsection{The $\eta_c\rightarrow \gamma\gamma$ decay width}

\begin{figure}[htb]
\centering
\includegraphics[width=0.40\textwidth]{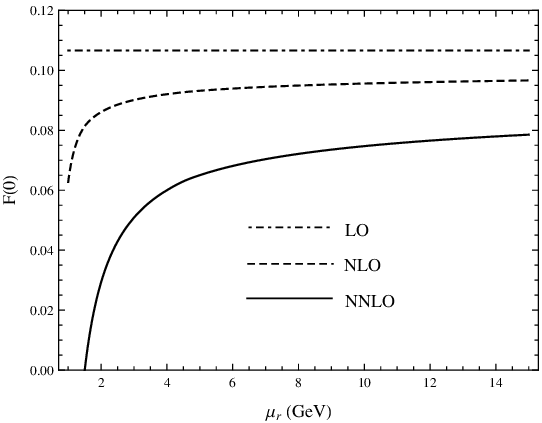}
\caption{The dependence of the renormalization scale $\mu_r$ for $F(0)$ using conventional scale setting. Dash-doted, dashed and solid lines stand for the QCD correction at LO, NLO, NNLO, respectively. The factorization scale is $\mu_\Lambda=1$ GeV.}
\label{F0scaledepe}
\end{figure}

In Fig.(\ref{F0scaledepe}), we present the dependence of the renormalization scale $\mu_r$ for $F(0)$ using conventional scale setting. The LO terms are free from strong interactions. Fig.(\ref{F0scaledepe}) shows that the NLO correction provides a moderate negative contribution to $F(0)$. In contrast, the NNLO correction gives an anomalous sizeable negative contribution, showing a strong dependence on the renormalization scale $\mu_r$. Thus, with the inclusion of the NNLO terms, the scale uncertainty and the convergence of perturbative series become worse. For example, the QCD corrections for $F(0)$ are:
\begin{eqnarray}
F(0)&=&0.1066-0.0438-0.2276=-0.1648, \\
F(0)&=&0.1066-0.0253-0.0815=-0.0001, \\
F(0)&=&0.1066-0.0165-0.0392=0.0509
\end{eqnarray}
for $\mu_r=1$ GeV, $m_c$ and $2m_c$, respectively. The predicted value of $F(0)$ is even negative for $\mu_r=m_c$. The scale uncertainty and the convergence of pQCD series get worse at small scales and improve at large scales. Thus, the quality of the convergence of the pQCD series depends on the choice of renormalization scale, and one cannot decide whether the poor convergence is an intrinsic property of the pQCD series, or is simply due to the improper choice of the renormalization scale.

By using the PMC method, due to the magnitude of the $n_f$-dependent term $\delta_{n_f,V}^{(2)}(\mu_r)$ is quite small, the scale-independent PMC NNLO conformal coefficient is close to the conventional NNLO coefficient, i.e., $\delta_{con,V}^{(2)}(\mu_r)=-49.41$ using the PMC and $\delta_V^{(2)}(\mu_r)=-38.84^{+4.88}_{-2.85}$ for $\mu_r\in[1$ GeV, $2m_c]$ using conventional scale setting for $\mu_\Lambda=1$ GeV. Thus the final value of the pQCD series for $F(0)$ is almost determined by the scale of the coupling $\alpha_s$. By using Eq.(\ref{PMCscaleV}), the PMC scale in the $V$ scheme is
\begin{eqnarray}
Q_\star=4.49\,m_c=6.74 \,\,\rm{GeV}
\end{eqnarray}
with $m_c=1.5$ GeV, for any choice of $\mu_r$. The PMC scale can be regarded as the effective momentum flow of the process. Thus, the effective momentum flow for the $\eta_c\rightarrow \gamma\gamma$ decay process is very different from the $c$-quark mass $m_c$ and is much larger than the conventional choice of $\mu_r=m_c$. The QCD prediction for the $F(0)$ for $\mu_\Lambda=1$ GeV using the PMC scale setting is:
\begin{eqnarray}
F(0)=0.1066-0.0123-0.0245=0.0698
\end{eqnarray}
for any choice of $\mu_r$. In this case, the NNLO QCD correction is greatly suppressed, and the convergence of pQCD series is greatly improved using the PMC scale setting compared with the conventional results. However, the convergence of PMC series is still poor due to the fact that the NNLO conformal coefficient is large. Thus, the intrinsic convergence of the pQCD series is notably slow at NNLO for the $\eta_c\rightarrow \gamma\gamma$ process, indicating the importance of uncalculated NNNLO QCD terms. It is interesting to find that if one assumes a large scale $\mu_r$ using the conventional scale-setting, the scale uncertainty and the convergence of pQCD series for $F(0)$ gets also improved, and the resulting pQCD predictions are close to the PMC results. In conclusion, a better choice of the renormalization scale for the $F(0)$ is given by a much larger value with respect to the conventional choice $\mu_r=m_c$.

The conventional predicted values for $F(0)$ are $-0.1648,\,\,-0.0001,\,\,0.0509$ for $\mu_r=1$ GeV, $m_c$ and $2m_c$, respectively. In contrast, the scale-independent predicted value for $F(0)$ is fixed to $0.0698$ by applying the PMC. The renormalization scale uncertainty is eliminated. In addition to the renormalization scale $\mu_r$ ambiguity, there are other sources of uncertainty, such as the NRQCD matrix element, the factorization scale $\mu_\Lambda$, the $c$-quark mass $m_c$ and the value of $\alpha_s(M_Z)$. The factorization scale $\mu_\Lambda$ uncertainty exists also for conformal theories, and it can be set by matching the perturbative prediction with the nonperturbative bound-state dynamics~\cite{Brodsky:2014yha}.

In the case of conventional scale setting for $\mu_r=m_c$, we obtain $F(0)=-0.0001,\,\,-0.0308,\,\,-0.0807$ for $\mu_\Lambda= 1$ GeV, $m_c$ and $2m_c$, respectively. The resulting values for $F(0)$ are negative for $\mu_\Lambda\in[1$ GeV, $2m_c]$ and are also plagued by large factorization scale uncertainty. After applying the PMC scale setting, we obtain for $F(0)$ the values $0.0698,\,\,0.0603,\,\,0.0464$ for $\mu_\Lambda= 1$ GeV, $m_c$ and $2m_c$, respectively. We observe that the $F(0)$ is positive for $\mu_\Lambda\in[1$ GeV, $2m_c]$ and is affected by a reasonable factorization scale $\mu_\Lambda$ uncertainty.

We also note that the dependence on the $c$-quark mass $m_c$ is larger when using conventional scale setting, i.e., $F(0)=-0.0241,\,\,-0.0001,\,\,0.0116$ for $m_c=1.4,\,1.5,$ and $1.6$ GeV, respectively. In contrast, the PMC prediction for $F(0)$ displays a more physically reasonable $c$-quark mass dependence, i.e., $F(0)=0.0799,\,\,0.0698,\,\,0.0614$ for $m_c=1.4,\,1.5,$ and $1.6$ GeV, respectively.

By using Eq.(\ref{EQ:F0}), the resulting $\eta_c\rightarrow \gamma\gamma$ decay width $\Gamma_{\eta_c\rightarrow\gamma\gamma}$ using the conventional scale setting is in complete disagreement with the PDG value $\Gamma_{\eta_c\rightarrow \gamma\gamma}=5.1\pm0.4$ keV. Even considering the worst case when all sources of errors lead to large uncertainties, one cannot eliminate this large discrepancy by simply using the conventional scale setting.

\begin{figure}[htb]
\centering
\includegraphics[width=0.40\textwidth]{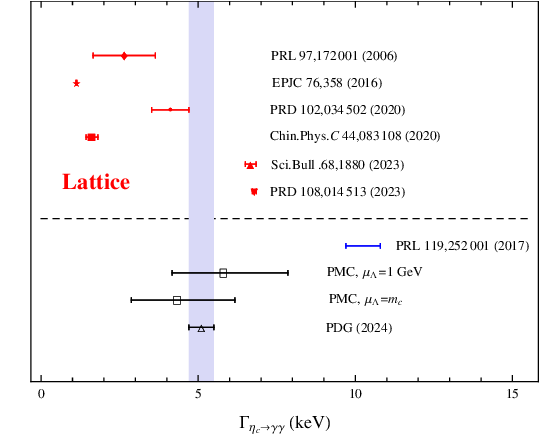}
\caption{A comparison of the PMC predictions for the decay width $\Gamma_{\eta_c\rightarrow\gamma\gamma}$ with the PDG value and the lattice QCD results. The NRQCD result in Ref.\cite{Feng:2017hlu} is also shown.}
\label{F0scaleConPMC}
\end{figure}

The values for the decay width $\Gamma_{\eta_c\rightarrow\gamma\gamma}$ obtained by using the PMC scale setting are:
\begin{eqnarray}
\Gamma_{\eta_c\rightarrow\gamma\gamma}&=&5.79_{-1.32-0.92-0.15}^{+1.79+1.00+0.15}\,\,\rm{keV} \,\,\rm{for}\,\, \mu_\Lambda= 1\,\, \rm{GeV}, \nonumber\\
\Gamma_{\eta_c\rightarrow\gamma\gamma}&=&4.32_{-1.05-0.98-0.16}^{+1.48+1.11+0.15}\,\,\rm{keV} \,\,\rm{for}\,\, \mu_\Lambda= m_c,
\end{eqnarray}
where the first error is determined by varying the $c$-quark mass $m_c\in[1.4,1.6]$ GeV, the second error is caused by the estimation of unknown higher-order contributions, and the third error is due to $\Delta\alpha_s(M_Z)=\pm0.0009$. A reliable estimation of unknown higher-order contributions for the scale-independent conformal series is often obtained~\cite{Du:2018dma} with the help of the Pad$\acute{e}$ approximation approach (PAA)~\cite{Basdevant:1972fe, Samuel:1992qg, Samuel:1995jc}. We also use the PAA method to estimate unknown higher-order contributions. Due to the large conformal coefficient for the decay width $\Gamma_{\eta_c\rightarrow\gamma\gamma}$, the pQCD convergence is still poor and the PMC predictions have relatively large errors. The PMC predictions obtained by using these two typical factorization scales, i.e. $\mu_\Lambda=1$ GeV and $m_c$ are in agreement with the PDG value $\Gamma_{\eta_c\rightarrow \gamma\gamma}=5.1\pm0.4$ keV. More explicitly, in Fig.(\ref{F0scaleConPMC}) we present a comparison of the PMC predictions for the decay width $\Gamma_{\eta_c\rightarrow\gamma\gamma}$ with the PDG value and the lattice QCD results. Both the LQCD results and the NRQCD result~\cite{Feng:2017hlu} cannot comprehensively explain the PDG value.

\subsection{The transition form factor}

In addition to the decay width $\Gamma_{\eta_c\rightarrow\gamma\gamma}$, NRQCD predictions for the transition form factor (TFF) $F(Q^2)$ also show severe inconsistencies with the experimental measurements. The NRQCD prediction including the NNLO QCD corrections~\cite{Feng:2015uha} fails to explain the \textit{BABAR} measurements over a wide range of values of the momentum transfer squared $Q^2$~\cite{BaBar:2010siw}. As a consequence of this recently arisen discrepancy, the applicability of the NRQCD approach has been considered questionable~\cite{Feng:2015uha}

\begin{figure}[htb]
\centering
\includegraphics[width=0.40\textwidth]{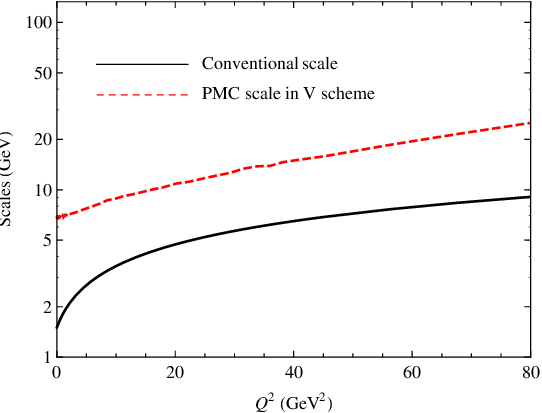}
\caption{The PMC scale for $F(Q^2)$ versus the momentum transfer squared $Q^2$. The conventional choice $\mu_r=\sqrt{Q^2+m^2_c}$ (solid line) is presented as a comparison. }
\label{FQscalePMC}
\end{figure}

Also for the case $F(Q^2)$, analogously to $F(0)$, the NNLO correction gives an anomalous sizeable negative contribution, and shows a strong dependence on the renormalization scale $\mu_r$. As for the $F(0)$ case, we use the same methodology for calculating $F(Q^2)$. Given that, the PMC-NNLO and the conventional-NNLO coefficients have similar magnitudes, the predicted value for $F(Q^2)$ is mostly entirely determined by the scale in $\alpha_s$ related to the method used to set it. We present the PMC scale of TFF $F(Q^2)$ versus $Q^2$ in Fig.(\ref{FQscalePMC}). We notice in Fig.(\ref{FQscalePMC}) that the PMC scale varies according to the momentum transfer squared $Q^2$. More importantly, scales show completely different behaviors when using conventional scale setting or PMC scale setting; the magnitude of the PMC scale is larger than the conventional choice $\mu_r=\sqrt{Q^2+m^2_c}$, especially, in the small $Q^2$ region.

\begin{figure}[htb]
\centering
\includegraphics[width=0.40\textwidth]{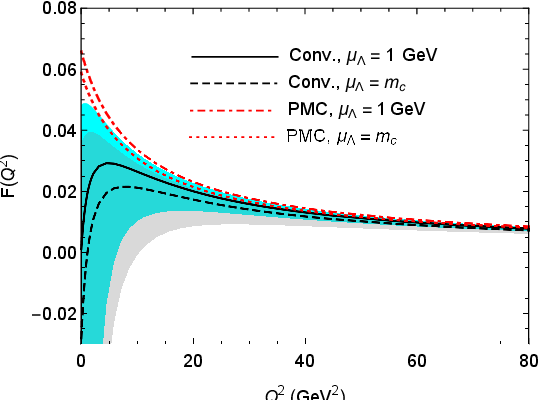}
\caption{The TFF $F(Q^2)$ versus the momentum transfer squared $Q^2$ using conventional and PMC scale settings, where the shaded bands show the conventional scale uncertainty for $\mu^2_r\in[(Q^2+m^2_c)/2,2(Q^2+m^2_c)]$. The PMC prediction (dashed line) is independent of the scale $\mu_r$. }
\label{FQscaleConPMC}
\end{figure}

In Fig.(\ref{FQscaleConPMC}), we present the TFF $F(Q^2)$ versus the momentum transfer squared $Q^2$ using conventional and PMC scale settings. The PMC prediction does not depend on the choice of the scale $\mu_r$, whereas the conventional prediction encounters a large scale $\mu_r$ uncertainty, especially in the low $Q^2$ region. In fact, in the low $Q^2$ region, the predicted value of $F(Q^2)$ significantly depends on the choice of the scale in $\alpha_s$, and thus the $F(Q^2)$ predictions are very different if using the conventional scale setting or the PMC scale setting. For the case of PMC scale setting, the large PMC scales will reduce significantly the NNLO correction term, and thus the predicted value for $F(Q^2)$ are larger than the conventional result. It is worth mentioning that the $F(Q^2)$ decreases monotonically while increasing $Q^2$ by using the PMC, whereas the $F(Q^2)$ rises and then drops while increasing $Q^2$ using the conventional scale setting.

In 2010, the \textit{BABAR} Collaboration has measured the ratio $|F(Q^2)/F(0)|$ over the range of $2$ GeV$^2<Q^2<50$ GeV$^2$. The measurements can be parameterized as $|F(Q^2)/F(0)|=1/(1+Q^2/\Lambda)$ with $\Lambda=8.5\pm0.6\pm0.7$ GeV$^2$~\cite{BaBar:2010siw}, and the ratio $|F(Q^2)/F(0)|$ decreases monotonically while increasing the $Q^2$ value. On the theoretical side, the nonperturbative matrix element cancels for the ratio $|F(Q^2)/F(0)|$. In the case of conventional scale setting, the anomalous rise-then-drop behavior for $|F(Q^2)/F(0)|$ fails to explain the \textit{BABAR} measurements. In addition, the ratio $|F(Q^2)/F(0)|$ using conventional scale setting encounters strong dependence on the renormalization scale $\mu_r$, the factorization scale $\mu_\Lambda$ and the $c$-quark mass $m_c$. Even allowing large uncertainties due to these sources of errors, the large discrepancy between the conventional results and the experimental measurements~\cite{BaBar:2010siw} cannot be eliminated.

\begin{figure}[htb]
\centering
\includegraphics[width=0.40\textwidth]{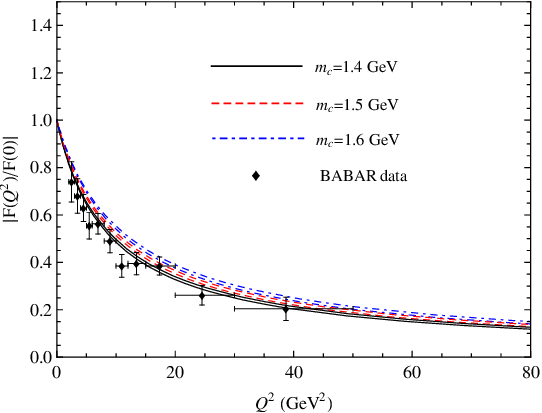}
\caption{The ratio $|F(Q^2)/F(0)|$ versus the momentum transfer squared $Q^2$ using PMC scale setting, where the lower (upper) line is for $\mu_\Lambda= 1$ GeV ($\mu_\Lambda=m_c$) for the same type of lines. The \textit{BABAR} data are shown as a comparison~\cite{BaBar:2010siw}. }
\label{FQF0FQOF0}
\end{figure}

In Fig.(\ref{FQF0FQOF0}), we present the ratio $|F(Q^2)/F(0)|$ versus the momentum transfer squared $Q^2$ using PMC scale setting for $\mu_\Lambda= 1$ GeV and $\mu_\Lambda=m_c$. It shows that the ratio $|F(Q^2)/F(0)|$ decreases monotonically when increasing the $Q^2$ value for the PMC scale setting case. PMC results successfully explain the \textit{BABAR} measurements~\cite{BaBar:2010siw} over a wide range of $Q^2$ within the uncertainties. The PMC predictions eliminate the renormalization scale uncertainty. The dependencies on the factorization scale $\mu_\Lambda$ and on the $c$-quark mass $m_c$ are greatly suppressed.

\section{Summary}
\label{sec:4}

In summary, according to the NRQCD, the NNLO pQCD predictions for the decay width $\Gamma_{\eta_c\rightarrow\gamma\gamma}$ and the transition form factor $F(Q^2)$ using conventional scale setting are affected by strong dependence on the renormalization scale $\mu_r$, as well as on the factorization scale $\mu_\Lambda$ and on the $c$-quark mass $m_c$. Even allowing large uncertainties for these error sources, the conventional pQCD results do not fit the precise experimental measurements. The ratio $|F(Q^2)/F(0)|$ does not involve any adjustable nonperturbative matrix element parameters, and the relativistic correction $\mathcal{O}(\alpha_sv^2)$ is negligible. These discrepancies have casted doubt on the applicability of NRQCD to charmonium processes.

However, as we have shown, the discrepancies with NRQCD are caused by the improper analysis of the highly scale-dependent fixed-order pQCD series. In contrast, by applying  PMC scale-setting, we observe that a precise renormalization scale-invariant pQCD series can be achieved. Moreover, the factorization scale $\mu_\Lambda$ and $c$ quark mass uncertainties are greatly suppressed. The poor pQCD convergence of the PMC series indicates the importance of uncalculated NNNLO QCD terms for this process. The PMC predictions for the decay width $\Gamma_{\eta_c\rightarrow\gamma\gamma}$ and the transition form factor agree with the experimental measurements within reasonable errors. The application of PMC renormalization scale-setting thus suggests a potential resolution to $\eta_c\rightarrow \gamma\gamma$ puzzle and supports the applicability of NRQCD to heavy quarkonium processes.

\hspace{1cm}

{\bf Acknowledgements}: The authors would like to thank Chao-Qin Luo for helpful discussions. This work was supported in part by the Natural Science Foundation of China under Grant No.12265011, No.12175025 and No.12347101; by the Project of Guizhou Provincial Department under Grant No.YQK[2023]016 and ZK[2023]141; by the Hunan Provincial Natural Science Foundation with Grant No.2024JJ3004, YueLuShan Center for Industrial Innovation (2024YCII0118); and by the Department of Energy Contract No.DE-AC02-76SF00515. SLAC-PUB-250127.

\end{document}